\documentclass[fleqn]{annalen}
\usepackage{graphics}
\begin{document}
\newcommand{\volume}{8}              
\newcommand{\xyear}{1999}            
\newcommand{\issue}{Spec. Issue}     
\newcommand{\recdate}{30 July 1999}  
\newcommand{\revdate}{dd.mm.yyyy}    
\newcommand{\revnum}{0}              
\newcommand{\accdate}{1 August 1999} 
\newcommand{\coeditor}{M. Schreiber} 
\newcommand{\firstpage}{65}       
\newcommand{\lastpage}{68}        
\setcounter{page}{\firstpage}        
\newcommand{\bm}[1]{\mbox{\boldmath $#1$\unboldmath}}
\newcommand{\keywords}{strongly correlated systems, 
                       non--linear effects,
                       fermions in reduced dimensions} 
\newcommand{\PACS}{71.10Pm, 72.20Ht, 42.65Ky}
\newcommand{\shorttitle}{A. Fechner et al., Local fields...} 
\title{Local fields in non--linear quantum transport}
\author{A. Fechner$^{1}$, G. Cuniberti$^{2}$, M. Sassetti$^{3}$, 
        and B. Kramer$^{1}$} 
\newcommand{\address}
  {$^{1}$ I. Inst. f\"ur Theor. Physik, 
          Universit\"at Hamburg,
          Jungiusstra\ss e 9,  
          D-20355 Hamburg \\
   $^{2}$ Max--Planck--Inst. f\"ur Physik komplexer Systeme,
          N\"othnitzerstra\ss e 38, 
          D-01187 Dresden \\
   $^{3}$ Dipartimento di Fisica, 
          INFM, 
          Universit\`a di Genova, 
          Via Dodecaneso 33,
          I-16146 Genova}
\newcommand{\email}{\tt fechner@physnet.uni-hamburg.de} 
\maketitle
\begin{abstract}
  We investigate the dynamical interplay between currents and
  electromagnetic fields in frequency-dependent transport through a
  single-channel quantum wire with an impurity potential in the
  presence of electron-electron interactions.  We introduce and
  discuss a formalism which allows a self-consistent treatment of
  currents and electromagnetic fields.
\end{abstract}

\section{Introduction}
\label{Introduction}
In order to describe frequency-dependent quantum transport
properties, it is necessary to take into account local fields.
However, since the microscopic charge and current distributions are
rapidly varying functions of space and time the solution of Maxwell's
equations is a complex task. Especially in the presence of non--linear
effects, a self-consistent theory of frequency and time dependent
transport is very challenging \cite{Keller}.

Here, we address this question for a simple but nonetheless
non-trivial system, namely a single-channel quantum wire of
interacting electrons with one potential barrier embedded in
three dimensions (3D). The interaction is taken into account using the
Luttinger model \cite{HaldaneJPC, Voit}. The system 
has a non--linear dc-current-voltage characteristic due to the
simultaneous presence of both, the electron-electron interaction and
the tunnel barrier \cite{KanePRB, CunibertiEPL}. 
We demonstrate explicitly the essential features of self--consistency 
for an infinite barrier. The connection between local and external field 
is worked out and generalized to the situation of a finite barrier.
The results of the
present considerations are the starting point for the more general
case, namely the non-linear, frequency dependent transport effects in
(1D) disordered systems containing many impurities.

\section{Local fields}
\label{Local-fields}

For a discussion of local-field effects one has to distinguish between
longitudinal electric fields, $ \nabla \times {\bm{E}}_{L} ({\bm{r}},
\omega) = 0$, and transverse electric fields, $ \nabla \cdot
{\bm{E}}_{T} ({\bm{r}}, \omega) = 0$. Using the microscopic Maxwell
equations and the continuity equation, they are connected with each
other and with the currents as well as in general an external field
\cite{Keller},
\begin{eqnarray}
\label{eq:EL}
{\bm{E}}_{\rm L} ({\bm{q}}, \omega) & = & {\bm{E}}_{\rm L,
ext} ({\bm{q}}, \omega)
      + \frac{1}{{\rm i} \epsilon_0 \omega} 
      {\bm{J}}_{\rm L} ({\bm{q}}, \omega)
\\
\label{eq:ET}
{\bm{E}}_{\rm T} ({\bm{q}}, \omega)
& = & {\bm{E}}_{\rm T, ext} ({\bm{q}}, \omega)
      + {\rm i} \mu_0 \omega g_0 ({\bm{q}}, \omega)
      {\bm{J}}_{\rm T} ({\bm{q}}, \omega),
\end{eqnarray}
where $g_0 ({\bm{q}}, \omega) = [|{\bm{q}} \, |^2 - (\omega/c)^2]^{-1}$.
The longitudinal and transverse currents are related to the total
current ${\bm{J}}$ by ${\bm{J}}_{\rm L} ({\bm{q}}, \omega) = {\bm{q}} \,
[{\bm{q}} \cdot {\bm{J}}({\bm{q}}, \omega)]/q^2$ and ${\bm{J}}_{\rm T}
({\bm{q}}, \omega) = {\bm{J}} ({\bm{q}}, \omega) - {\bm{J}}_{\rm L}
({\bm{q}}, \omega)$. In order to evaluate the local fields one has to
close eqs.~(\ref{eq:EL}) and (\ref{eq:ET}) by a  transport equation
which connects currents and local fields in terms of a microscopic
theory. In a linear system, this would be the Kubo
theory. In the general case, the non-linear relation between the
current and the local field has to be used. 

\section{The model}
\label{Model}

In order to apply Maxwell's equations, which
incorporate naturally Coulomb interaction, the quantum wire has to be
considered as embedded in 3D. This is achieved by assuming a
confinement perpendicular to the, say, $x$-direction and projecting the
relevant operators onto the eigenfunctions of the confinement
Hamiltonian. We use $\Phi_{nk}(\bm{r}) = e^{{\rm i} k x}
\varphi_{n}(\bm{R})/\sqrt{L}$ as a complete set of functions where
$k$ is the wavenumber along the direction of the wire, $L$ the
system length along $x$, $\varphi_n (\bm{R })$ the confinement
wave function ($\bm{R}=(y,z)$), and $n$ the corresponding quantum
numbers.

The Hamiltonian consists of the kinetic energy, the
electron-electron interaction, the confinement energy, the coupling to
the scalar electric potential $V(\bm{r}, t)$, and the potential
barrier of height $U_{\rm b}$ at $x_{\rm b}$, 
$H = H_{\rm kin} + H_{\rm conf} + H_{\rm int} + H_{\rm em} + H_{\rm b}$, 
with
\begin{eqnarray}
H_{\rm int} & = & \sum_{n, l, s, p} 
                  \int \int {\rm d}x {\rm d}x'
                  \Psi_n^{\dagger}(x) \Psi_l(x)
                  U_{nlsp} (x - x')
                  \Psi_s^{\dagger}(x') \Psi_p(x')
\\
\label{eq:Hem}
H_{\rm em} & = & e \int {\rm d}{\bm{r}} \rho (\bm{r}) V({\bm{r}},t) ,
\\
H_{\rm b} & = & \int {\rm d}{\bm{r}} \rho (\bm{r}) U_{\rm b}({\bm{r}}) ,
\end{eqnarray}
where, 
\begin{equation}
U_{n l s p} (x - x') = \int_S \int_{S'} 
                       {\rm d}^{2} R\, {\rm d}^{2} R' \, 
                       \varphi_n^* (\bm{R }) \varphi_l (\bm{R })
                       U_{\rm ee} (\bm{r} - \bm{r}\,')
                       \varphi_s^* (\bm{R }') \varphi_p (\bm{R }')
\end{equation}
is the projected interaction potential,
$\Psi_n(x)$ the fermion field of the $n$-th band,
\begin{equation}
  \label{eq:charge}
 \rho (\bm{r})=\sum_{n,l} \varphi ^{*}_{n}(\bm{R })
\varphi _{l}(\bm{R }) \Psi^{\dagger}_n (x) \Psi_l (x) ,
\end{equation}
the charge density. The current is related to
$\rho ({\bm{r}})$ via the 3D continuity equation.

The scalar electric potential in eq.~(\ref{eq:Hem}) is the local
potential. In general, the local vector potential should also be
included because even for a longitudinal external source, a transverse
field is generated. However, the amplitude of the
induced transverse field is of the order $(v_{\rm F}/c)^2$,
thus negligible \cite{FechnerI}, and 
eq.~(\ref{eq:ET}) does not contribute. 

The above general expressions are needed in order to treat 
especially quantum wires
with more than one subband. Here, we
restrict to only one subband, thus $n=l=s=p=0$.

The dispersion relation of the collective
excitations obtained by using the Luttinger model
\cite{HaldaneJPC,Voit} is
$\omega(q) = |q| v_{\rm F} [1 + U(q) / \pi \hbar v_{\rm F}]^{1/2}$,
where $q \equiv q_x$.
It reflects the Fourier transform of the interaction
potential, $U(q) \equiv U_{0000} (q)$
\cite{CunibertiPRB}.

\section{Some simple examples}
 \label{examples}

Neglecting currents and induced fields
perpendicular to the wire, eq.~(\ref{eq:EL}) can
be rewritten in terms of the local field projected 
onto the lowest subband,
\begin{equation}
\label{eq:EL1D}
E(q,\omega )= E_{\rm ext} (q, \omega) 
- \frac{i q^2}{e^2 \omega} U(q) J(q) , 
\end{equation}
where $J(q)$ is the $x$ component of the current.

The clean Luttinger wire was already discussed in
\cite{CunibertiPRB}. The frequency--dependent 
conductivity is in this case
\begin{equation}
\sigma(q, \omega) = \frac{{\rm i} e^2 v_{\rm F}}{\pi \hbar}
                    \frac{\omega}{\omega^2 - \omega^2(q)} .      
\end{equation}
One can show that
the current of the non--interacting system
driven by the local field is equivalent to
the current through the interacting system in response 
to the external field,
$ E(q, \omega) \sigma_0 (q, \omega) 
= E_{\rm ext} (q, \omega) \sigma (q, \omega)$,
where $\sigma_0 (q, \omega)$ corresponds to the conductivity of
the non--interacting wire.
In the following we will see that this remains true
in the presence of a tunnel barrier of infinite strength where
the non--linearity does not play any role.

In the presence of a single potential barrier,
assumed to be uniform over the cross--section of the wire, 
$U_{\rm b} ({\bm{r}}) = U_{\rm b} \delta(x - x_{\rm b})$,
a relation can be derived 
which links the current at different positions \cite{SassettiPRB}. 
In wave number space,
\begin{equation}
\label{eq:linearcurrent}
I(q, \omega) = I_0(q, \omega) 
             + r (q, \omega) [I_{\rm b} (\omega) - I_{0, \rm{b}}(\omega)] 
\end{equation}
where $I_0(q, \omega)$ represents the current in the clean system,
$I_{\rm b} (\omega)$ is the current at the barrier, 
and the ratio of conductivities without barrier at the 
two positions considered is
$r(x, \omega) = \sigma(x-x_{\rm b}, \omega) / \sigma(x=0, \omega)$.
The total current is non--linear due to the presence of 
$I_{\rm b}$ \cite{FechnerII}.

In order to demonstrate how self--consistency
affects the electric field inside the wire, 
we start with an infinite barrier, $I_b=0$, 
decoupling the system into two separate wires.
In this case, dc--transport is forbidden but
the displacement contribution to the current is finite.

We evaluate the linear current in eq.~(\ref{eq:linearcurrent}) in 
terms of the local field by
using the conductivity of the non--interacting system. 
Inserting this 
into eq.~(\ref{eq:EL1D}), gives a
self--consistent equation for the local field.
The solution is
\begin{equation}
\label{eq:locfieldres}
E(q, \omega) 
= \frac{\sigma(q, \omega)}{\sigma_0(q, \omega)} E_{\rm ext}(q, \omega) 
- \frac{\sigma(q, \omega)}{\sigma_0(q, \omega)} \,
  e^{{\rm i} q x_{\rm b}}
  \frac{\int dq \, \sigma(-q, \omega) e^{-{\rm i} q x_{\rm b}}
                E_{\rm ext}(q, \omega) }
       {\sigma(x=0, \omega)} .
\end{equation}
It is important to note that an equivalent result would have been obtained
by driving the current with the external field using the conductivity
of the interacting system in complete analogy with the single clean
wire mentioned above. This is consistent with the 
common textbook knowledge \cite{Mahan}
that without
a non--linearity the Coulomb interaction is 
automatically taken into account self--consistently in the
Luttinger model.

The second term in eq.~(\ref{eq:locfieldres}) 
represents the contribution due to 
the barrier. In order
to show its influence on the local field 
we consider $\omega \rightarrow 0$.
For zero--range interactions it generates a delta peak
at the position of the barrier which is broadened when
increasing the range of the interaction.

For a finite barrier, $I_b$ is
highly non--linear due to the 
presence of the interaction \cite{KanePRB}.
It is then easy to show from eq.~(\ref{eq:linearcurrent}) that 
the two cases, (i) interacting system 
driven by external field and (ii) non--interacting system 
driven by local field are
no longer equivalent.
In order to take properly into account the combined effect
of the interaction,
the non--linearity as well as the local fields,  
the current at the barrier has to be driven by 
the local field but here incorporating also the interaction. 
\section{Summary}
\label{Summary}
We investigated self-consistency in time-dependent transport through a
non--linear system, in particular through a quantum wire with a tunnel
barrier. For an infinite barrier, we obtained that by
renormalising the electric field the current is given by the
conductivity of the non-interacting system. In 
the non-linear case, one has to consider both,
the renormalized field and the interaction.

\vspace*{0.25cm} \baselineskip=10pt{\small \noindent This work was
  supported by the EU within the TMR program, INFM via
  PRA(QTMD)97, Cofinanziamento (MURST)98 and the Deutsche
  Forschungsgemeinschaft via the SFB 508 of the Universit\"at Hamburg}
%
%
%
%
%
%
%
%
%
%
%
%

\end{document}